\documentclass[runningheads]{llncs}

\usepackage{epsfig}  
\usepackage{amssymb} 
\usepackage{latexsym} 

\begin{document}
\setcounter{page}{47}
\title{CLPGUI: a generic graphical user interface for constraint logic programming over finite domains}
\titlerunning{CLPGUI: a generic graphical user interface for CLP(FD)}
\author{Fran{\c{c}}ois Fages}
\authorrunning{F. Fages}
\institute{Projet Contraintes, INRIA-Rocquencourt,\\
           BP105, 78153 Le Chesnay Cedex, France, \\
           \email{Francois.Fages@inria.fr}}

\maketitle

\addtocounter{footnote}{1}
\footnotetext{In Alexandre Tessier (Ed), proceedings of the 12th International Workshop on Logic Programming Environments (WLPE 2002), July 2002, Copenhagen, Denmark.\\Proceedings of WLPE 2002: \texttt{http://xxx.lanl.gov/html/cs/0207052} (CoRR)}

\begin{abstract} 
CLPGUI is a graphical user interface for visualizing and interacting
with constraint logic programs over finite domains. In CLPGUI, the
user can control the execution of a CLP program through several views
of constraints, of finite domain variables and of the search tree.
CLPGUI is intended to be used both for teaching purposes, and for
debugging and improving complex programs of realworld scale.  It is
based on a client-server architecture for connecting the CLP process
to a Java-based GUI process. Communication by message passing provides
an open architecture which facilitates the reuse of graphical
components and the porting to different constraint programming
systems.  Arbitrary constraints and goals can be posted incrementally
from the GUI.
We propose several dynamic 2D and 3D visualizations of the search tree
and of the evolution of finite domain variables.  We argue that the 3D
representation of search trees proposed in this paper provides the most
appropriate visualization of large search trees.  We describe the
current implementation of the annotations and of the interactive
execution model in GNU-Prolog, and report some evaluation results.
\end{abstract} 
 

\section{Introduction}

Several tools for visualizing the execution of constraint programs
have been developed in the last few years.
These tools have been found very useful 
for debugging and improving
constraint programs, and for teaching constraint programming.
One can distinguish:
\begin{itemize}
\item {\em post-mortem visualization} tools, these tools are used after execution
of the program, the program is annotated with specifications of the information to trace.
This approach is implemented for example in the CHIP or CIAO systems, it allows using a wide variety of
viewers, including both application oriented tools \cite{SCDFNT00discipl},
and generic tools, for visualizing the search tree \cite{CH00discipl,SA00discipl}, finite domain variables \cite{CH00discipl2},
or constraint propagation \cite{SABB00discipl}.
\item {\em dynamic visualization} tools, these tools are connected to the constraint programming interpreter
and realize a visualization on-line, possibly with animations \cite{GB00discipl}. This approach is implemented in 
Grace tool \cite{Meier95cp} for finite domains visualization, and in OPL studio \cite{BGP01wlpe}
for search tree and constraint propagation visualization.
\item {\em dynamic visualization and control} tools which allow interaction
with a CLP process through different visualizations. One example is the Oz-Explorer system \cite{Schulte97cp}
where it is possible to jump to any previously encountered state by simply clicking on a node of the search tree,
and restart computation from that state. User-guided search is implemented in Oz-Explorer using the first-class computation spaces of Oz.
Recomputation is used to trade space for time in Oz-Explorer, and similarly in OPL studio \cite{BGP01wlpe},
the state restoration mechanisms in tree search are described in \cite{CHK01cp}.
\end{itemize}

In this paper we propose to push forwards these ideas
towards an open architecture for connecting a CLP process
with dynamic visualization and control tools. 
To our knowledge, the visualization of search trees in three dimensions has not been much investigated.
We argue that the 3D representation of search trees proposed in this paper 
provides the most appropriate visualization of large search trees.

\begin{figure}[htb]
\begin{center}
\epsfig{file=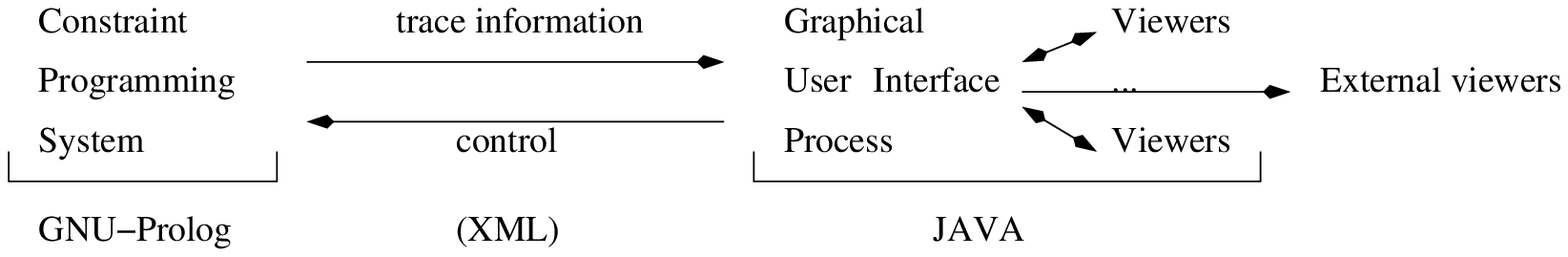, width=12.5cm}
\end{center}   
\caption{Information flow for dynamic visualization in CLPGUI.}\label{schema}
\end{figure}

Our ambition is not to realize an {\em ad hoc} tool limited to a particular
constraint programming system, namely GNU-Prolog \cite{Diaz01gprolog}, but a generic tool which can be ported
to other constraint programming systems. 
The main reason for this is that a wide variety of viewers can be useful for debugging
constraint programs and it is possible in this way to share developments.
Our approach relies on the generic trace format which is defined in the OaDymPPac consortium \cite{OADymPPaC}
for post-mortem analysis.
We propose to extend this format with a similar generic format for control, and to use these formats
for connecting on-line the CLP process to the GUI process.
Our implementation of CLPGUI \cite{Fages01clpgui} in GNU-Prolog and Java is depicted in Figure \ref{schema}.

In Section \ref{console} of this paper, we describe the client-server architecture of CLPGUI and the user console
which facilitates the establishment of a connection with a CLP server, and controls the execution of the CLP program.
Section \ref{annotations} describes annotations that can be added to a CLP program,
in order to give an external name to finite domain variables,
to create buttons for posting constraints or goals from the CLPGUI console,
and to define the level of granularity of the search tree which will be visualized.
Section \ref{finitedomains} presents a dynamic 3D viewer for visualizing the evolution of finite domains
variables over time.
Section \ref{searchtree} describes the representation of the explored search space
by partial CSLD derivation trees, presents different visualizations with 2D and 3D viewers,
and discusses their use in CLPGUI.
Then in Section \ref{02/implementation} we discuss the implementation of the interactive execution model
and of the annotations, using the global variables of GNU-Prolog.
Finally we provide some evaluation results, and conclude on the current limitations of the system and 
on some perspectives for future work.

\section{Client-Server Architecture}\label{console}

\begin{figure}[htb]
\begin{center}
\epsfig{file=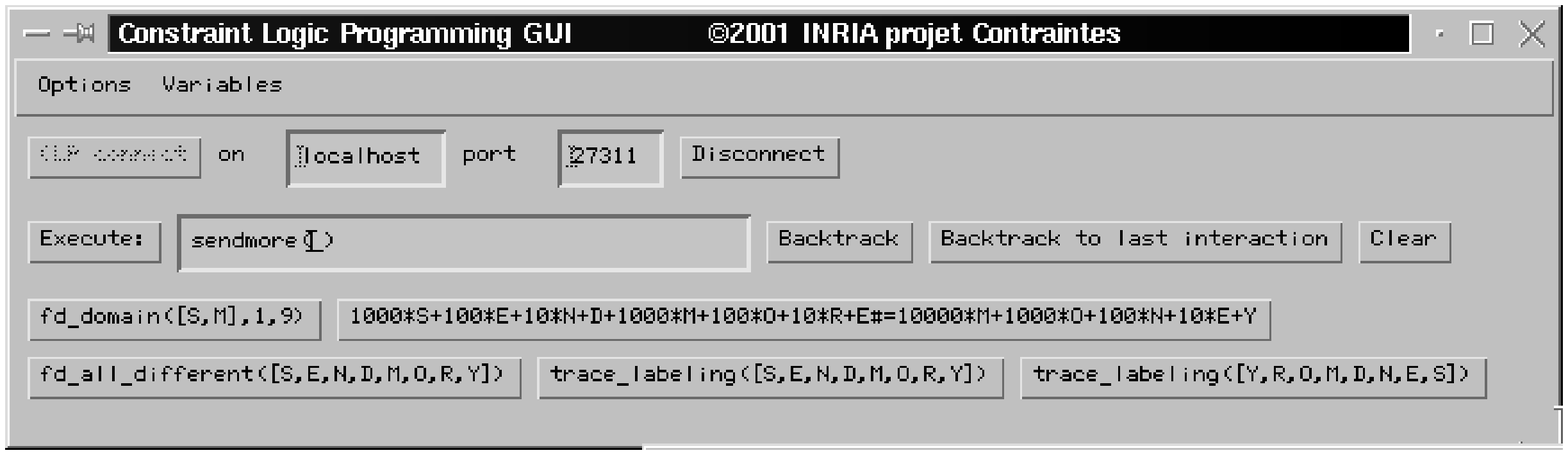, width=12cm}
\caption{CLPGUI console.}\label{consoleCLPGUI}
\end{center}   
\end{figure}
The graphical user interface of CLPGUI is a Java application
connected by sockets as a client to a server which executes CLP goals.
Both processes can run on different machines and communicate over the network.
This has been experienced with CLPGUI for visualizing the execution of CLP programs on a Workbench of
Virtual Reality.
The choice of the Java language for implementing the GUI is motivated by several reasons:
\begin{itemize}
\item its object-orientation, all 3D viewers presented in the following sections inherit from a single
class for moving and projecting 3D figures;
\item the encapsulation of events handling, that is preponderant in dynamic visualization;
\item the threaded execution, which is mandatory for implementing communication with the CLP process;
\item its wide availability.
\end{itemize}
For efficiency reasons, we did not use the Java-3D library for the viewers presented in this paper,
as they can directly benefit from {\em ad hoc} optimizations that speed-up their incremental display.
Nevertheless the architecture can support the use of the powerful Java-3D library
for developing complex application-oriented viewers.

During initialization, the CLP server starts an interpreter of the command lines
received on the socket. The command lines may contain any CLP goal.
The GUI Java client opens a graphical console such as the one in Figure \ref{consoleCLPGUI}.
That console is used for establishing a connection to the CLP server,
and for posting constraints or executing CLP goals. 

The CLP program may contain annotations for creating buttons for some
constraints or for some Prolog goals to execute in an interactive manner.
These buttons for posting constraints or Prolog goals then appear at the bottom in the CLPGUI console, see Figure \ref{consoleCLPGUI}.
A click on the button posts the constraint or executes the goal associated to the button.
Other arbitrary goals can be executed by entering them in a text field.
In addition, one button called ``backtrack'' continues the execution of the current goal up to the next success,
or, if there are no more success, returns to the state of the previous interaction.
Another button called ``backtrack interaction'' forces backtracking to the state of the previous interaction.
The menu bar of this console contains menus to select and activate the viewers of the search tree or of
the finite domain variables.

\section{Annotations}\label{annotations}

The CLP program may contain annotations for giving an external name to CLP(FD) variables,
for creating buttons for posting constraints or goals from the CLPGUI console,
and for marking the goals to visualize in the search tree.
The following predicates are part of the annotation library:
\begin{itemize}
\item {\tt varnames([V1,...,VN],[name1,...,nameN])} and {\tt varnames([V1,...,Vn])}
give an external name to a list of CLP variables.
These external names are used in the graphical user interface and for the communication by sockets.
\item {\tt button(goal)} creates a button in the GUI console for executing a goal or for posting a constraint.
\item {\tt bagof\_buttons(goal, call)} creates a bag of buttons for each successful instance of the second argument.
\item {\tt trace\_search(goal)} executes the goal and traces the execution of that goal, by creating nodes in the search tree.
The goals and constraints posted from the graphical console are always traced.
\item {\tt show\_domains} updates the visualization of the current state of FD variables.
\end{itemize}

Annotations provide a simple mechanism for encapsulating communications towards the GUI process \cite{CH00discipl}.
The advantages of this abstraction are:
\begin{itemize}
\item the flexibility for defining different levels of granularity 
concerning the information to visualize.
\item the easiness for making interactive already existing programs,
\item the portability of the GUI to other constraint programming systems,
as all communications with the GUI are encapsulated in the implementation of annotation predicates.
\end{itemize}
The limitations of annotations are well-known in standard programming environments:
they may be difficult to maintain in large programs. In that case, one solution is to automatically
generate annotations with a graphical editor of the program source, where spy points and trace options can be specified.
Nevertheless, one peculiarity of constraint logic programming is the conciseness of programs.
CLP(FD) programs for solving combinatorial optimization problems on real-size data
may compute with a huge amount of constraints and variables, but the program source for handling
constraints and defining complex search strategy usually remains relatively concise.
Therefore in this context, the proposed annotations appear as a satisfactory solution.

In many CLP systems however, the heuristic labeling procedures are built-in,
and cannot be precisely traced with a simple annotation.
In these cases, the annotations have to rely on the tracing facilities of the CLP system
in order to extract, and communicate to the GUI, the relevant information.
A simpler solution is to program the labeling heuristics in the host language,
for making available in the host language the information coming from the constraint solvers
that is relevant to the search heuristics.
In that case, the effect of the search strategy can be visualized at different levels of granularity.
In its simplest form, a predicate for tracing a labeling procedure can be defined 
with a {\tt trace\_search} annotation as follows:
{\small
\begin{verbatim}
trace_labeling([]).
trace_labeling([X|L]):- trace_search(fd_labeling(X)), trace_labeling(L).
\end{verbatim}} It is worth noting that even if the search strategy is implemented with a meta-interpreter, 
or constraint posting is combined with labeling,
the relevant goals of the execution can still be traced with annotations.

The same difficulty arises for tracing internal constraint propagation steps.
This is not possible without access to the wakening events of the constraint solver.
The annotations for tracing constraint propagation steps have thus to rely
on the tracing facilities of the solver in order to extract and communicate constraint wakening events.

\begin{example}\label{send.pl}
The following annotated GNU-Prolog program solves the well known
SEND+MORE=MONEY puzzle in an interactive manner,
by creating buttons for posting the constraints and for trying two labeling goals in this example:
{\small
\begin{verbatim}
sendmore(L):-
         L=[S,E,N,D,M,O,R,Y],
         varnames(L,['S','E','N','D','M','O','R','Y']),
         fd_domain(L,0,9),
         button(fd_domain([S,M],1,9)),
         button(1000*S+100*E+10*N+D + 1000*M+100*O+10*R+E 
                #= 10000*M+1000*O+100*N+10*E+Y),
         button(fd_all_different(L)),
         button(trace_labeling(L)),
         reverse(L,L2),
         button(trace_labeling(L2)).
\end{verbatim}
}
This program generates the console in Figure \ref{consoleCLPGUI}.
The evolution of the finite domain variables over time, after the posting of constraints and of the
first labeling goal, is depicted in Figure \ref{domainsCLPGUI}.
The visualization of the search tree for obtaining all solutions under the first
labeling goal, and then under the second labeling goal executed after a backtracking command,
is depicted in Figure \ref{sendtreeCLPGUI}.
Under the first ordering, the labeling is deterministic.
Under the second ordering, few backtracking steps occur on variable {\tt Y} when searching for other solutions.
Note that other labeling heuristics can be tried directly from the console.
On such pedagogical examples, the advantage of immediately visualizing the effect
of posting a constraint or trying a labeling, is clear for teaching purposes.
\end{example}

\begin{figure}[htb]
\begin{center}
epsfig{file=sendtreeCLPGUI.ps, width=8cm, height=5cm}
\end{center}
\caption{Search trees with two labeling orderings in the puzzle SEND+MORE=MONEY.}\label{sendtreeCLPGUI}
\end{figure}

\begin{example}\label{queens.pl}
The following program solves the N queens problem
by creating buttons for posting the constraints ({\tt safe} predicate)
and labeling goals for each variable ({\tt fd\_labeling} predicate) and for all variables ({\tt trace\_labeling} predicate).
{\small
\begin{verbatim}
queens(N,L):-
        length(L,N),
        varnames(L),
        fd_domain(L,1,N),
        button(safe(L)),
        bagof_buttons(fd_labeling(X),member(X,L)), 
        button(trace_labeling(L)).    
\end{verbatim}
}
Three visualizations of the search tree for the 8 queens problem are depicted in Figures \ref{arbre2DCLPGUI}, \ref{arbre3DCLPGUI}
and \ref{treemapCLPGUI}.
\end{example}

\section{3D Views of Finite Domain Variables}\label{finitedomains}
\begin{figure}[htb]
\begin{center}
epsfig{file=domainsCLPGUI.ps, width=12cm, height=8cm}
\end{center}
\caption{Dynamic 3D view of finite domain variables in the puzzle SEND+MORE=MONEY.}\label{domainsCLPGUI}
\end{figure}

The evolution of finite domain variables is visualized
in a three dimensional graph variable-domain-time, as proposed in the VIFID/TRIFID tool \cite{SCH99pac}.
Here the visualization is dynamic, the Java process reads the stream of finite domains information
and paints the figure in an incremental manner. 
Domains are depicted by their size on the vertical axis, see Figure \ref{domainsCLPGUI}.
According to options, that can be set in the CLP program or in the GUI,
only the size, the interval or the complete domain of variables is visualized.
But in any case only the sizes of the domains are memorized, therefore the extra information is
lost when the figure is repaint. 
The time axis traces the interactions (i.e.~the posting of constraints in the example), and the execution of traced goals
(i.e.~the labeling in the example). This view shows that the posting of constraints instantiate variables $S,M,O$ 
and that the first labeling step on variable {\tt E} in fact instantiates
all variables by constraint propagation.
An option determines whether backtracked states are traced or erased.

The figure can be moved, zoomed and rotated.
For efficiency reasons, the rotations are limited to a quadrant of a sphere
which is not a real limitation for the user.
In this way the visible faces are efficiently determined and
the figure can be drawn incrementally.

Extra information on variables and executed goals can be obtained by moving the mouse
on the position of a variable or on a time position.

The 3D dynamic view of finite domain variables evolution is very useful for teaching constraint programming.
The effect of constraints is immediately seen and many strategies can be tried step by step.
On large set of variables, the 3D view of domains can still be useful to get a view of the pruning power of
different constraint modelings, and of the efficiency of different search heuristics, by comparing
the general shape of domain reductions.

\section{2D and 3D Views of the Search Tree}\label{searchtree}
\subsection{Partial CSLD derivation trees}

\begin{figure}[htb]
\begin{center}
epsfig{file=arbre2DCLPGUI.ps, width=12cm, height=8cm}
\end{center}
\caption{2D view of the search tree in the 8-queens problem.}\label{arbre2DCLPGUI}
\end{figure}
 
\begin{figure}[htb]
\begin{center}
epsfig{file=arbre3DCLPGUI.ps, width=12cm, height=8cm}
\end{center}
\caption{3D of the search tree in the 8-queens problem.}\label{arbre3DCLPGUI}
\end{figure}
 
\begin{figure}[htb]
\begin{center}
epsfig{file=treemapCLPGUI.ps, width=12cm, height=8cm}
\end{center}
\caption{Treemap representation obtained by rotation of the 3D view.}\label{treemapCLPGUI}
\end{figure}
 
The search tree considered in CLPGUI is a labeled tree defined as follows:
\begin{itemize}
\item a node is introduced for each call to a traced goal (called a {\em call node}), 
and for each success to a traced goal (called a {\em success node}),
\item the label of a call node is the called goal,
\item the label of a success node is the list of named variables with their value,
\item the arcs correspond to the operational CLP transitions.
\end{itemize}

This tree is a subtree of the CSLD derivation tree \cite{JM94jlp}.
It is thus a quite natural representation of the search tree for describing CLP program execution.
A branch represents a conjunction, and the different successors of a node represents a disjunction.
A success node may have several successors if
there is an untraced non-deterministic goal which is executed after the success,
and before the next call to a traced goal.
This is the main reason why success nodes are introduced in partial CSLD trees.
In this way, the non-determinism due to untraced goals cannot be confused with the
non-determinism of traced goals.

One disadvantage of CSLD trees is that in the case of deterministic programs
they are threadlike and thus space consuming in their standard representation.
AND-OR trees provide a more compact representation, as the threadlike parts of the CSLD tree
are compacted in the successors of a single AND-node. 
For this reason, in the context of logic programs where most predicates are deterministic,
AND-OR trees, and their variant AORTA diagrams which 
indicate the status of resolution of the goals,
have been preferred \cite{EB88jlp}.
Nevertheless in the context of constraint logic programming over finite domains,
the situation is quite different. The search tree to visualize is usually focused
on the labeling predicates, or more generally on the branching procedure,
which is highly non-deterministic (at least during debugging). The representation of the deterministic part
of the search tree with threadlike structures provides an immediate visualization 
of the pruning power of constraints.

A naive solution for tracing constraint propagation steps in this approach is to
add deterministic nodes for tracing constraint wakening events.
For space limitation reasons, it is preferable however to aggregate constraint propagation information
to the nodes of the search tree. This is proposed in the ``Christmas trees'' of OPL studio \cite{BGP01wlpe}.

For search engines not based on backtracking, it is worth noting that
a partial CSLD derivation tree can still provide a valid representation of the explored search space,
as long as the explored states can be defined by their relation
to some ancestor states. A formalization of an interactive constraint solver 
by transformations of CSLD derivation trees was done in \cite{FFS95iclp}.

\subsection{Visualization}
Once the search tree is formally defined, it can still be visualized in many ways, 
and in some cases it can be interesting to use several visualizations at the same time.
We have currently implemented several two-dimensional and three-dimensional viewers,
but many more representations could be imagined and fruitfully used.

In all the following representations, the labels of the nodes are visualized when the mouse is moved on them,
and an option makes all nodes visible.
The successes are materialized by a red cross.
Each view can be moved, zoomed and rotated.

Figure \ref{sendtreeCLPGUI} uses a standard 2D representation of the search tree in a fixed width.
Figure \ref{arbre2DCLPGUI} uses a dynamic 2D representation of the tree with a fixed spacing between leaves.
This representation of the tree can be drawn incrementally and is thus appropriate
for the dynamic visualization of large trees.

To our knowledge, the 3D visualization of search trees has not been much investigated.
Figure \ref{arbre3DCLPGUI} shows a somewhat original 3D representation of the search tree with alternating planes of successors.
One advantage of this 3D representation is that it is relatively compact,
it helps visualizing rather large trees by playing with rotations, see Figure \ref{bridgeCLPGUI} for another example.
Our experience is that the 3D view is the most appropriate view to apprehend the shape
of large search trees.

It is interesting to note that one obtains a treemap representation of the tree
by rotation of the 3D alternate tree up to its vertical projection, as done in Figure \ref{treemapCLPGUI}.
Treemap representations (with colors for aggregating information)
are known to be particularly efficient to represent very large data \cite{Schneiderman92tog}
and to visualize complex phenomenons such as correlations, patterns or symmetries.
                      
The interaction allowed in these views to restore a state is currently 
limited to user-guided backtracking and recomputation.
The automatic recomputation of any state of the tree as described in \cite{BGP01wlpe,CHK01cp,Schulte97cp}
is currently not implemented.

\subsection{Branch and bound optimization}\label{optimization}

\begin{figure}[htb]
\begin{center}
epsfig{file=bridge2.ps, width=12cm, height=8cm}
\end{center}
\caption{3D view of the search tree in the bridge problem (3905 call nodes).}\label{bridgeCLPGUI}
\end{figure}

The branch and bound procedure is widely used in constraint programming
to solve optimization problems. Branch and bound optimization develops search trees
in two parts. The first part corresponds to the enumeration of solutions with 
decreasing costs (for minimization problems). The second part exhausts the search space
to show that there does not exist a better solution than the last solution found.
The second part of the search tree constitutes the proof of optimality.

Figure \ref{bridgeCLPGUI} shows the search tree for the bridge problem \cite{VH89mit},
a medium size job-shop scheduling problem. The first descent corresponds to the search of the first solution
of cost 108. It contains 78 call nodes. 
The second descent (after some hesitation) corresponds to the search of the optimal solution of cost 104.
It contains 99 call nodes. The bottom part of the tree contains 3728 call nodes and corresponds to the proof of optimality.
The 3D view is the most appropriate view of this search tree. It can be moved and rotated without difficulty.

\section{Implementation}\label{02/implementation}

\subsection{Interactive Execution Model for CLP}

The interactive execution model of the CLP process derives from 
a more general model for adding and removing constraints and goals described
in \cite{FFS95iclp}. In CLPGUI, constraints and goals can only be added to the current goal,
the removing of constraints or goals occurs by backtracking.
It is therefore possible on a success of the current goal:
\begin{itemize}
\item 
to add constraints or any goals to the current goal
and continue resolution,
\item to backtrack to the next success (command ``backtrack'' of Section \ref{console}),
\item or to backtrack to the last interaction (command ``backtrack interaction'').
\end{itemize}
It is worth noting that such a top level is in fact very appropriate for standard Prolog systems,
where the capability of adding goals on a success of the current goal, and continue resolution,
is usually missing.

Our current implementation uses the global variables of GNU-Prolog \cite{Diaz01gprolog}
to memorize global information, such as input and output sockets, variable names, 
and information used for backtracking.
Global variables make it possible to avoid adding parameters to many predicates and lead to a simple
implementation of annotations.

\subsection{Communication messages}

In this section we describe the communication messages which are transmitted between
the CLP process and the GUI process.
The CLP process produces the trace information specified by the annotations in the CLP program,
or asked from the GUI.
The messages emitted from the CLP to the GUI are the following:
\begin{itemize}
\item {\tt <variables ...>} sends the list of FD variable names
\item {\tt <button G>} asks the GUI to create a button for posting the constraint or goal {\tt G}
\item {\tt <undo button G>} indicates backtracking on the creation of a button for {\tt G}
\item {\tt <node G>} traces a call to goal {\tt G}
\item {\tt <undo node G>} traces backtracking on the call to {\tt G}
\item {\tt <child G>} traces a success to {\tt G}
\item {\tt <undo child G>} traces backtracking on the success to {\tt G}
\item {\tt <undo goal G>} indicates backtracking on the call to goal {\tt G}
\item {\tt <domainSizes ...>} sends the domain sizes of FD variables
\item {\tt <domainIntervals ...>} sends the current intervals of FD variables
\item {\tt <domainValues ...>} sends the current finite domains of FD variables
\item {\tt <undo domainValues>, <undo domainIntervals>, <undo domainSizes>} warns the GUI that 
the finite domain variables are updated by backtracking.
\item {\tt <success>} indicates that the current derivation is a success
\item {\tt <clear>} indicates return to top level.
\end{itemize}

In the other direction, the messages emitted from the GUI to the CLP process are the following:
\begin{itemize}
\item {\tt <showSize>, <showInterval>, <showValues>} sets the information on finite domains that need be sent
\item {\tt <execute G>} asks to post constraint {\tt G} or execute goal {\tt G}
\item {\tt <backtrack>} asks backtracking to the next success
\item {\tt <backtrackInteraction>} forces backtracking to the last interaction
\item {\tt <clear>} asks to abort the current execution.
\end{itemize}

The portability of CLPGUI to a new constraint programming system is determined by the ability of 
the constraint programming system to produce and interpret these communication messages.
The messages of the first list are produced by the predicates of
annotation library described in Section \ref{annotations}.
The messages in the second list are interpreted by the interactive execution model described in the previous section.

\section{Evaluation}

Our experience of using CLPGUI for teaching constraint programming has been very positive.
The dynamic visualization of CLP programs really speeds-up the process of learning
the basic concepts of domain filtering, constraint propagation and search trees.
CLPGUI has also been fruitfully used to visualize the search tree of CLP(R) programs.

On real-size data, CLPGUI has shown satisfactory performance figures.
We report in this section the timings on a Pentium III 600 MHz processor under Linux.
GNU-Prolog solves the bridge problem mentioned in Section \ref{optimization}
in 100 ms, including the proof of optimality. 
The solving together with the visualization of the search tree with 3905 call nodes 
takes 470 ms with CLPGUI.
This overhead is due to the communication of messages by sockets.
The overhead was reduced from 2600 ms to 470 ms by optimizing socket calls and by
using simple data compression techniques for communication.
Moreover, the drawing of the tree is immediate and the figure can be moved and rotated
without difficulty.

\section{Conclusion and future work}\label{conclusion}

We have described an open architecture for visualizing and controlling the execution of 
constraint logic programs. Communication by sockets between the CLP process and the GUI process
has proved efficient enough for dynamic visualization and interactions.
An important reduction of the overhead was obtained by optimizing socket calls and by
using simple data compression techniques for communication.

CLPGUI supports the use of different viewers.
Our experience has shown that 
a somewhat original 3D visualization of the search tree proposed in the paper, is often the preferred view
to apprehend the shape of large search trees, as it is very compact.
More work is needed however to parametrize the different viewers
and invent novel visualizations of complex data.
In this respect the flexibility of the architecture makes it possible to connect CLPGUI
to external generic viewers, or to use powerful libraries like Java 3D
to develop application-oriented viewers.

The most obvious limitation of our current implementation of CLPGUI is 
the absence of connection to a system for tracing constraint propagation,
simply because such a tracer does not exist yet for GNU-Prolog.
Nevertheless a generic trace format for finite domain constraint solvers has been defined in 
the OaDymPpac consortium \cite{OADymPPaC},
and we plan to use this format in future versions of CLPGUI.

Finally, CLPGUI is not a visual programming tool as far as the capabilities of defining goals from the GUI
are extremely rudimentary. Nevertheless the proposed architecture can support this kind of
extension by adding the capability to define constraints and goals graphically, that is certainly worth investigating.

\subsubsection*{Acknowledgements.}{\small 
I would like to thank all members of the OaDymPpac RNTL project 
of the French Ministry of Research, and especially 
Abderrahmane Aggoun, Thomas Baudel, Pierre Deransart, Jean-Daniel Fekete, Ludovic Langevine and Mohammad Gonhiem
for interesting discussions on this topic.
I am grateful also to Anupam Agarwal for optimizing communication by sockets,
and to Jean-Michel Leconte for his preliminary work along these lines
on the workbench of virtual reality at INRIA.}

%

\end{document}